\documentclass[12pt]{article}
\pdfoutput=1

\usepackage{draft} 
\usepackage{hyperref}
\usepackage{graphicx,color,subfig}
\usepackage{cite}
\usepackage{mciteplus}
\usepackage{skak}
\DeclareFontFamily{OT1}{pzc}{}
\DeclareFontShape{OT1}{pzc}{m}{it}{<-> s * [1.10] pzcmi7t}{}
\DeclareMathAlphabet{\mathpzc}{OT1}{pzc}{m}{it}

\numberwithin{equation}{section}

\def\0{{(0)}}
\def\1{{(1)}}
\def\2{{(2)}}

\def\<{\langle }
\def\>{\rangle }

\def\eads${H$_3$}

\def\be#1\ee{\begin{align}#1\end{align}}

\begin{document}

\begin{titlepage}

\begin{center}

\hfill \\
\hfill \\
\vskip 1cm

\title{Light-ray Operators and the BMS Algebra}

\author{Clay C\'ordova and Shu-Heng Shao}

\address{
School of Natural Sciences, Institute for Advanced Study, \\Princeton, NJ 08540, USA\\
}

\email{claycordova@ias.edu, shao@ias.edu}

\end{center}

\vspace{2.0cm}

\begin{abstract}
\noindent
We study light-ray operators constructed from the energy-momentum tensor in $d$-dimensional Lorentzian conformal field theory.  These include in particular the average null energy operator.     
The commutators of parallel light-ray operators  on a codimension one light-sheet form an infinite-dimensional algebra.  
We determine this light-ray algebra and find that the $d$-dimensional (generalized) BMS algebra, including both the supertranslation and the superrotation, is a subalgebra.  
We verify this algebra in correlation functions of free scalar field theory.  We also determine the infinite-dimensional algebra of light-ray operators built from non-abelian spin-one conserved currents.

\end{abstract}

\vfill

\end{titlepage}

\eject

\tableofcontents

\section{Introduction}

In this paper we study a local version of the standard Poincar\'e symmetry algebra.  We consider non-local operators that are supported along a light-ray and can be expressed as integrals of conserved currents. One such light-ray integral is the average null energy (ANEC) operator 
\begin{equation}
\mathcal{E}\equiv \int_{-\infty}^{\infty}dx^-  T_{--}(x)~.  \label{anecdef}
\end{equation}
In addition, we also consider the light-ray operators
\begin{equation}
\mathcal{K}\equiv \int_{-\infty}^{\infty}dx^-  x^{-}T_{--}(x)~, \hspace{.5in}\mathcal{N}_{A}\equiv \int_{-\infty}^{\infty}dx^-  T_{-A}(x)~.  \label{otherlightray}
\end{equation}
Our main result is to determine the algebra obeyed by such light-ray operators using the elementary constraints of causality, unitarity, and Ward identities of Poincar\'e symmetries.  The algebra of these light-ray operators is therefore universal and exists in any unitary conformal field theory.\footnote{In our derivation, conformal symmetry enters through unitarity bounds on local operators.  It would be interesting to see if the algebra we find also holds in more general quantum field theories.}  In particular we show that the (generalized)  
Bondi-van der Burg-Metzner-Sachs (BMS) algebra of \cite{Bondi:1962px,Sachs:1962wk,Sachs:1962zza} is a subalgebra of smeared versions of these light-ray operators. We verify that this algebra is consistent with conformal Ward identities in Appendix \ref{app:ward}. We also discuss analogs for internal global symmetries and verify the light-ray algebra in the special case of free scalar field theory.

It is instructive to compare our results to a more standard analysis of symmetries in quantum field theory.  In general, a continuous global symmetry is associated with a conserved current local operator $j_{\mu}(x)$. The associated charge operator $Q$ is a non-local operator supported on a codimension one manifold $\Sigma$
\begin{align}
Q=  \oint_{\Sigma} d\Sigma^\mu  j_\mu(x)\,.
\end{align}
For instance, in the most elementary application $\Sigma$ is taken to be a spatial slice.  Current conservation implies that we can continuously deform $\Sigma$ (provided we do not cross any charged local operators) while leaving correlation functions invariant.  Thus, the charge $Q$ is a topological operator, a fact which is instrumental in deriving some of the most powerful consequences of symmetries such as Ward identities. (see e.g. \cite{Gaiotto:2014kfa})

To place the light-ray operators on a similar footing we further smear them on a codimension one light-sheet ${\cal S}:~x^+=0$ in Minkowski spacetime with metric $ds^2= -dx^+dx^- +|dx^\perp|^2$.  For example, the smeared version of the ANEC operator \eqref{anecdef} is given by 
\begin{align}\label{smear}
{\cal T}(f)  \equiv  \oint_{\cal S}d{\cal S}^\mu \, f(x^\perp) T_{\mu -}(x) = \int d^{d-2}x^\perp \, f(x^\perp )\, \mathcal{E}(x^{\perp})\,,
\end{align}
where above $f(x^{\perp})$ is an arbitrary function of the transverse coordinates on $\cal S$.  This is a local version of the translation charge  along the $x^-$ direction since in the special case of constant $f$, \eqref{smear} reduces to $P_-$.  However, there is an important distinction between these more general non-local operators ${\cal T}(f)$ and the Poincar\'e charge operator $P_-$.  Since the underlying current (i.e.\ $f(x^\perp) T_{\mu -}$, for ${\cal T}(f)$) is not conserved, the non-local operator ${\cal T}(f)$ is not a topological operator.   In particular, the definition of ${\cal T}(f)$ depends on the details of the codimension one manifold where it is supported.  

Despite these differences, we will still see that these non-local operators satisfy an interesting infinite-dimensional algebra if they all lie on the same light-sheet $\cal S$.  For instance, when interpreted as part of the BMS algebra ${\cal T}(f)$ plays the role of a supertranslation.  In particular, this means that the $\mathcal{T}(f)$ operators commute with each other.  In terms of the original light-ray operators, this is equivalent to the statement that parallel ANEC operators on the same light-sheet commute
\begin{equation}
[\mathcal{E}(x_{1}^{\perp}),\mathcal{E}(x_{2}^{\perp})]=0~.
\end{equation} 
The same conclusion has also been reached in \cite{Casini:2017roe} (see also \cite{Balakrishnan:2017bjg}).

Our work touches on diverse subjects, including the null energy condition, quantum information theory, asymptotic symmetries, and soft theorems.  Let us expand on these connections below.  

 The ANEC operator $\mathcal{E}$ has been central to many recent developments in quantum field theory.  One of its most important features is that it is non-negative, i.e.\ it has a positive expectation value in any state:  $\langle \Psi| \mathcal{E}|\Psi \rangle \geq 0.$  This result was established using conformal field theory methods in \cite{Hartman:2016lgu}, and using information theoretic techniques in \cite{Faulkner:2016mzt}.  In the conformal collider setup of \cite{Hofman:2008ar}, this positivity is applied to constrain conformal field theory data such as OPE coefficients.  In our context these constraints mean that the physical representation of the algebra of light-ray operators on the Hilbert space enjoys positivity conditions analogous to unitarity constraints.  Further aspects of light-ray operators are discussed in detail in \cite{Kravchuk:2018htv, KKSDZ}.

The connection between the light-ray operators such as $\mathcal{T}(f)$ and quantum information theory has been explored in  \cite{Casini:2017roe}.  There, the difference between the modular Hamiltonian of regions having future horizon lying on the light-sheet $\mathcal{S}$ and the modular Hamiltonian of their complementary regions were shown to be given by linear combinations of smeared versions of \eqref{anecdef} and the first term of \eqref{otherlightray}.  Moreover in that context, the algebra of light-ray operators $\int dx^-(x^{-})^{n}T_{--}$ was determined.  We apply the same techniques to fix the algebra of operators  \eqref{anecdef} and \eqref{otherlightray} below.

The BMS algebra has recently made an extensive appearance in the study of soft graviton physics and its connection to asymptotic symmetries \cite{Strominger:2013jfa,He:2014laa}.  See \cite{Strominger:2017zoo} and references therein.  In that context the BMS generators are interpreted as adding soft (zero momentum) gravitons to the external states, and the resulting BMS symmetry of the S-matrix has as a consequence the soft graviton theorems \cite{Weinberg:1965nx}.  Connections between modular Hamiltonians and soft theorems have also been explored in \cite{Kapec:2016aqd}.

Although we will not explore this relationship in detail here, we observe that the light-ray operators are naturally related to asymptotic symmetries.  Indeed, since the algebra we find is constructed from the energy-momentum tensor it can be interpreted as the algebra of local modifications of the metric on the light-sheet.  By applying a conformal transformation, we can move this light-sheet to future null infinity and the BMS algebra of light-ray operators manifests as generators of the asymptotic symmetries.  Moreover, such operators are clearly connected to the insertion of soft gravitons since they are given by integrals of the energy-momentum tensor.  In particular, the supertranslation generator $\mathcal{T}(f)$ has vanishing null momentum $P_{-}$.  Therefore when acting on states of fixed $P_{-}$ it produces other degenerate states of the same null momentum, and it is tempting to speculate that this is related to the soft degeneracy \cite{Strominger:2013jfa} and memory effect described in \cite{Strominger:2014pwa,Pasterski:2015tva}.  We leave a more detailed investigation of this direction to future work.

\section{The BMS Algebra}

We will work in $d$-dimensional Minkowski spacetime $\mathbb{R}^{d-1,1}$ with metric $ds^2 = -dx^+dx^- +dx^Adx^A$, where 
$x^\pm  =x^0  \pm x^1$ are the lightcone coordinates and $x^A$ with $A=2,3,\cdots d-1$ are the transverse coordinates.  The transverse coordinates $\{x^A\}$ will often be collectively denoted by $x^\perp$.  
Let $\cal S$ be a codimension one light-sheet defined as 
\begin{align}
{\cal S}:~x^+=0\,.
\end{align}
We will denote the transverse $\mathbb{R}^{d-2}$  to the 01-plane by $\cal M$, i.e.\ ${\cal M}:~x^+ =x^-=0$.    The BMS algebra will be defined on $\cal M $.

  In the context of asymptotic symmetry and soft theorems, $\cal M$ is usually taken to be the celestial sphere $S^{d-2}$ at  the future (or past) null infinity in Minkowski spacetime $\mathbb{R}^{d-1,1}$ (see, for example, \cite{Strominger:2017zoo}).  This configuration is related to the above by  a conformal transformation \cite{Hofman:2008ar}:
  \begin{align}
  x'^{+} = -{1\over x^+}\,,~~~~x'^- = x^-   -  {|x^\perp|^2 \over x^+}\,,~~~~x'^A = { x^A \over x^+}\,.
  \end{align}
     For simplicity, we will work in the conformal frame where ${\cal M}= \mathbb{R}^{d-2}$ with a flat metric.

The  BMS algebra \cite{Bondi:1962px,Sachs:1962wk,Sachs:1962zza} is an infinite-dimensional algebra generated by the supertranlsations  and the superrotations.  The supertranslation, denoted as ${\cal T}(f),$ is generated by a scalar function $f(x^\perp)$ on $\cal M$.  
The superrotation, denoted as ${\cal R}(Y^A)$, is generated by  a vector field $Y^A(x^\perp)$ on $\cal M$.   Both operators are linear in their arguments
\begin{equation}
\mathcal{T}(f_{1}+f_{2})=\mathcal{T}(f_{1})+\mathcal{T}(f_{2})~, \hspace{.5in}\mathcal{R}(Y_{1}^{A}+Y_{2}^{A})=\mathcal{R}(Y_{1}^{A})+\mathcal{R}(Y_{2}^{A})~.
\end{equation}
The commutation relations in $d$ dimensions are \cite{Barnich:2011ct}
\begin{align}\label{BMS}
\begin{split}
&[{\cal T}(f_{1}) , {\cal T} (f_{2}) ]=0\,,\\
&[{\cal T}(f) , {\cal R}(Y^A) ] = i {\cal T}(g)\,, ~~~~~~~~~~~g = \frac {1}{d-2} f \partial_A Y^A -Y^A \partial_A f\,,\\
&[{\cal R}(Y_1^A) , {\cal R}(Y_2^A) ]
= i {\cal R}(Y_3^A) \,,~~~~~~~Y_3^A=  Y_1^B \partial_B Y_2^A -Y_2^B \partial_B Y_1^A\,.
\end{split}
\end{align}
In the standard BMS algebra, $Y^A(x^\perp)$ is restricted to be a globally well-defined conformal Killing vector.  The more general algebra with $Y^A(x^\perp)$ being any smooth vector field on $\cal M$ is called the generalized BMS algebra \cite{Campiglia:2014yka} (see also \cite{Banks:2003vp,Barnich:2009se,Barnich:2010eb,Barnich:2011ct,Barnich:2011mi}). 
In  Section \ref{sec:BMSsub} we will  reproduce the generalized BMS algebra from the commutators of light-ray operators.

\section{The Algebra of Light-ray Operators}\label{sec:alg}

\subsection{Light-ray Operators}

In this section we determine the algebra of parallel light-ray operators placed on the light-sheet ${\cal S}:~x^+=0$.  All operators we consider will be extended along the $x^-$ direction.

The operators we consider are expressed as integrals of the energy-momentum tensor:
 \begin{eqnarray}
  {\cal E}(x^\perp)  &\equiv&  \int _{-\infty} ^\infty dx^- T_{--}(x^- , x^+=0 , x^\perp) ~, \\
 {\cal K} (x^\perp)& \equiv& \int _{-\infty} ^\infty dx^- \, x^- \, T_{--}(x^- , x^+=0 , x^\perp)~,\\
   {\cal N}_A(x^\perp) &\equiv&\int _{-\infty} ^\infty dx^- \,  T_{-A}(x^- , x^+=0 , x^\perp)~.
 \end{eqnarray}
 In particular $\mathcal{E}(x^\perp)$ is the ANEC operator. 
 
  These light-ray operators are the local versions of some of the  charge operators of the Poincar\'e algebra.  If we further integrate them in the transverse directions with a constant profile, we obtain the generators of the Poincar\'e algebra:
 \begin{equation}\label{poincare}
 P_- = \int d^{d-2}x^\perp {\cal E}(x^\perp)~, \hspace{.35in}
 J_{01} = \int d^{d-2}x^\perp \, {\cal K}(x^\perp) ~, \hspace{.35in}
   P_A = \int d^{d-2}x^\perp {\cal N}_A(x^\perp)~,
 \end{equation}
where $P_-$ is the translation along the $x^-$ direction, $J_{01}$ is the boost on the $01$-plane, and $P_A$ is the translation along the $x^A$ direction.\footnote{  The boost  current is usually defined as
$x_0 T_{\mu1} -x_1 T_{\mu0}$, but can be written as 
$   x^- T_{\mu -}$ on the light-sheet $\cal S$ where $x^+=0$.}  The only nonzero commutator between the global  charges $P_-,J_{01},P_A$ is $[J_{01},P_-] = - i P_-$.

In the following we will compute the commutators of these parallel  light-ray operators ${\cal E}(x^\perp), {\cal K}(x^\perp), {\cal N}_A(x^\perp)$ on $\cal S$ separated by their transverse coordinates $x^\perp$.
We will determine the light-ray commutators  following the arguments in  \cite{Casini:2017roe}.  The  argument is based on:
\begin{itemize}
\item Microcausality:  Space-like separated operators commute with each other. 
\item Unitarity: All local operators transform in unitary representations of the conformal group $SO(d,2)$.  Such representations are classified in \cite{Mack:1975je,doi:10.1063/1.1705183,Minwalla:1997ka}. 
\item Ward Identities:  The global charges $P_-,J_{01},P_A$ implement the Poincar\'e transformations on operators.\footnote{Although we only use the Ward identities of Poincar\'{e} generators in our derivation, we verify that our results are also consistent with the Ward identities of conformal symmetry in Appendix \ref{app:ward}.
}
\end{itemize}

In addition to these basic assumptions which are valid in every unitary conformal field theory we also require a minimality hypothesis regarding the general properties of non-local operators supported on light-rays.  We elevate it to an additional assumption:\footnote{We thank P. Kravchuk and J. Maldacena for discussions on this point.}
\begin{itemize}
\item Closure: Given two parallel light-ray operators on $\mathcal{S}$ that can each be expressed as the integral of a conserved current local operator, their commutator can also be expressed as the integral of a local operator. 
\end{itemize}
Note that by comparison, the operator product expansion (OPE) of such parallel light-ray operators in general contains light-ray operators that cannot be written as the integral of local operators \cite{Hofman:2008ar, KKSDZ}.  It would be interesting to justify our hypothesis perhaps through a more detailed investigation of the OPE.

In \cite{Casini:2017roe}, the commutators involving ${\cal E}(x^\perp)$ and ${\cal K}(x^\perp)$ were determined using the assumptions above.  We review their derivation below and subsequently generalize to commutators involving ${\cal N}_A(x^\perp)$. 

Throughout, we will organize our derivation using the following two quantum numbers: the scaling dimension $\Delta$ and the boost charge on the $01$-plane $J$.   We will fix our convention for $J$ to be such that the coordinate $x^-$ has $J=-1$.  Their difference $\Delta-J$ is usually called the twist.  We will encounter candidate terms in the commutator which can be written as integrals of local operators with integral scaling dimension and boost charge.  If we assume our theory is generic and interacting, the only such operators are conserved spin one currents for global symmetries and the energy-momentum tensor.  This anticipates our conclusion that only integrals of such operators can appear in the light-ray algebra. In fact we will see that  $\mathcal{E}, \mathcal{K}, \mathcal{N}_{A}$ form a closed algebra.  One intuitive way to understand this is that after integrating over transverse coordinates we must recover the commutators of conformal symmetry and the appearance of other operators in the light-ray algebra would generally spoil this requirement.

\subsection{Computations of the Commutators}

\subsection*{$[{\cal E},{\cal E}]$}

Let us start with the commutator between two ANEC operators on $\cal S$:
\begin{align}\label{EE}
[{\cal E}(x_1^\perp) , {\cal E}(x_2^\perp)]
= \left[\int_{-\infty}^\infty dx_1^- \,  T_{--}(x_1^- ,x_1^+=0 ,x_1^\perp)  \,,\, \int_{-\infty}^\infty dx_2^- \,  T_{--}(x_2^- , x_2^+ = 0 ,x_2^\perp) \right] 
\,.
\end{align}
The Lorentzian separation between  the two energy-momentum tensor operators is $|x_{12}^\perp|^2$.  It follows that the two integrands are space-like separated as long as $x_{12}^\perp \neq0$. Hence the commutator only has support at $x_1^\perp = x_2^\perp$, i.e.\ there can  only be transverse contact terms on the right-hand side of \eqref{EE}.  Therefore the commutator takes the form
\begin{align}\label{RHS}
[{\cal E}(x_1^\perp) , {\cal E}(x_2^\perp)]
= \delta^{(d-2)} (x_{12}^\perp)  \, L(x_2^\perp)  + \partial_A \delta^{(d-2)} (x_{12}^\perp ) \,L^A (x_2^\perp) +\cdots\,,
\end{align}
where $\cdots$ represent terms with more derivatives on the transverse delta function and $L$ and $L^A$ are some non-local operators.    

Let us first show that $L=0$.  We perform the $x_1^\perp$ transverse integral on both sides to obtain:
\begin{align}
[P_- , {\cal E}(x_2^\perp) ] = L(x_2^\perp)\,,
\end{align}
Since the ANEC operator ${\cal E}$ is invariant under translation in the $x^-$ direction, the Ward identity for translation implies that the right-hand side must vanish, i.e.\ $L=0$.

What about the non-local operator $L^A$?  By our assumption of closure, $L^A$ can be written as an integral of a local operator $\Phi^A(x)$:
\begin{align}
L^A(x_2^\perp) = \int_{-\infty}^\infty dx^- \, g(x^-)\,\Phi^A(x^- ,x^+=0,x_2^\perp)\,, \label{ladef}
\end{align}
where $g(x^-)$ is some profile function.  With this ansatz, we can show that there is no unitary local operator $\Phi^{A}$ that can appear on the right-hand side of \eqref{ladef}.

To show this, let us compute the twist on the left-hand side of \eqref{EE}.  The scaling dimension of the ANEC operator is $\Delta= d-1$, while its boost charge is $J= +1$. Hence the twist on the left-hand side is $\Delta-J  =2 d-4$.  On the right-hand side, the transverse delta function $\partial_A\delta^{(d-2)}(x_{12}^\perp)$ has $\Delta= d-1$ and $J=0$, while the profile function $g(x^-)$ and $x^-$ do not contribute to the twist. It follows that the local operator $\Phi^A$ has twist $d-3$.  In a generic CFT the only operators of integral scaling dimension are spin-one currents and the energy-momentum tensor, and neither has twist $d-3$, thus implying that $\Phi^{A}$ must vanish.

More technically, in any CFT unitarity implies that local operators transforming in symmetric traceless vector representations (with $J>0$), the twist must be greater or equal to $d-2$
\begin{align}\label{twistgap}
\Delta- J  \ge d-2\,.
\end{align}
Equality is saturated by conserved currents.\footnote{More generally, there are local operators in conformal field theory transforming in Lorentz representations where unitarity permits smaller twist, e.g.\ a scalar operator. (An explicit presentation of such representations for $d=3,4,5,6$ can be found in Section 2.5 of \cite{Minwalla:1997ka}.)  We would like to argue that, even if they have appropriate integral scaling dimensions, such operators do not occur in the light-ray algebra.  

In $d=3$ the only such operators are Lorentz scalars and spinors, and the latter clearly do not appear in the algebra.  Moreover scalars also do not occur since the light-ray integrand must carry some Lorentz indices to match the boost charge.  In $d=4$ there are a priori more possibilities, however using the results of \cite{Cordova:2017dhq} and Bose symmetry we can restrict our attention to Lorentz scalars and two-forms.  As above, only the two-form is potentially relevant and could in principle appear in the commutator $[\mathcal{N}_{A}, \mathcal{N}_{B}]$ below.  However the two-form can only have integral scaling dimension below the twist gap if it is an exactly free Maxwell field.  For simplicity we assume that no such fields are present, though we do not expect them to appear in the light-ray algebra.  We anticipate that similar ideas can exclude the appearance of exotic representations in the light-ray algebra in $d>4$.}
Thus the candidate operator $\Phi^{A}$ violates the unitarity bound \eqref{twistgap} and $L^A=0$.  The terms in \eqref{RHS} with more derivatives of the transverse delta function are similarly excluded.

We conclude that parallel ANEC operators commute
\begin{align}\label{EE2}
[{\cal E}(x_1^\perp) , {\cal E}(x_2^\perp) ]=0\,.
\end{align}

\subsection*{$[{\cal K}, {\cal E}]$}

By causality, the commutator $[{\cal K}, {\cal E}]$ takes the form
\begin{align}
&[{\cal K}(x_1^\perp) , {\cal E}(x_2^\perp)]
= \delta^{(d-2)} (x_{12}^\perp)  \, L(x_2^\perp) 
\,,\label{KE}
\end{align}
where $L$ is some non-local operator, and terms with derivatives of $\delta^{(d-2)}(x_{12}^\perp)$ are ruled out by the same twist gap argument.  Below we will compute $L$. 

By integrating \eqref{KE} over $x_1^\perp$ we obtain the commutator $\left[ J_{01} , {\cal E}(x_2^\perp)\right],$  where $J_{01}$ is the boost generator on the $01$-plane. Since the ANEC operator has boost charge +1, this commutator must take the form
\begin{align}
\left[ J_{01} , {\cal E}(x_2^\perp)\right]  = - i {\cal E}(x_2^\perp)\,.
\end{align}
Incidentally, if we further integrate in $x_2^\perp$, we find the expected commutation relation between a boost and a translation in the Poincar\'e algebra: $[J_{01}, P_- ]  = -i P_-$.  Hence, we conclude that
\begin{align}\label{KE2}
[{\cal K}(x_1^\perp) , {\cal E}(x_2^\perp)]
=-i \delta^{(d-2)}(x_{12}^\perp)  \, {\cal E}(x_2^\perp)\,.
\end{align}

\subsection*{$[{\cal K} , {\cal K}]$}

Again by causality and the twist gap argument, the commutator takes the form
\begin{align}
&[{\cal K}(x_1^\perp) , {\cal K}(x_2^\perp)]=i \delta^{(d-2)} (x_{12}^\perp)  \, L(x_2^\perp)  
\,,
\label{KK}
\end{align}
where $L$ is some non-local operator. Integrating both sides and enforcing the fact that $\mathcal{K}$ is neutral under boosts in the 01-plane we conclude that $L=0$ i.e.\
\begin{align}\label{KK2}
[{\cal K}  (x_1^\perp)  , {\cal K}(x_2^\perp) ] =0\,.
\end{align}

Thus far we have reviewed the derivation in \cite{Casini:2017roe} of the commutation relations  \eqref{EE2}, \eqref{KE2}, \eqref{KK2} involving $\cal K$ and the ANEC operator $\cal E$.  We now move on to compute the commutators involving ${\cal N}_A$.

\subsection*{$[ {\cal N}_A,{\cal E} ]$}

The commutator is constrained by causality and unitarity to take the following general form
\begin{align}\label{NE}
&[{\cal N}_A(x_1^\perp) , {\cal E}(x_2^\perp)]
= i\delta^{(d-2)} (x_{12}^\perp) \, L_A(x_2^\perp)  +i \partial_B\delta^{(d-2)} (x_{12}^\perp) \, L_{AB}(x_2^\perp)\,,
\end{align}
where $L_A$ and $L_{AB}$ are some non-local operators.  Note that compared to the previous cases, the twist on the left-hand side is increased by 1. It follows that the term proportional to $\partial_B\delta^{(d-2)} (x_{12}^\perp)$ now survives the twist gap bound, while higher derivative terms are excluded.

Integrating $x_1^\perp$ on both sides of \eqref{NE}, gives the translation $P_A$ along the $x^A$ direction. The Ward identity implies that
\begin{align}
\left[ P_A  , {\cal E}(x_2^\perp) \right]  = -i  \partial_A {\cal E}(x_2^\perp)~,
\end{align}
and therefore
\begin{align}
L_A (x_2^\perp ) =  -  \partial_A {\cal E}(x_2^\perp) =- \int_{-\infty}^\infty dx_2^-  \, \partial_A T_{--}(x_2)\,.
\end{align}

To determine the operator $L_{AB}(x_{12}^\perp)$ in \eqref{NE}, we integrate both sides over $x_2^\perp$.   Since  ${\cal N}_A=\int dx^-T_{-A}$ is invariant under the translation along $x^-$ the result vanishes. It follows that
\begin{align}\label{inhom1}
 \partial_A {\cal E}(x_1^\perp) =\partial_B L_{AB}(x_1^\perp) ~.
\end{align}
Let us write
\begin{equation}
L_{AB}=\delta_{AB}\mathcal{E}+\hat{L}_{AB}~,
\end{equation}
so that the \eqref{inhom1} implies the following equation for the light-ray operator $\hat{L}_{AB}$:
\begin{equation}\label{hatl1}
\partial_{B}\hat{L}_{AB}=0~.
\end{equation}

We claim that the above equation implies that $\hat{L}_{AB}$ vanishes.  To see this we write $\hat{L}_{AB}$ as the integral of a local operator
\begin{equation}
\hat{L}_{AB}=\int_{-\infty}^{\infty}dx^{-}g(x^{-}) \mathcal{O}_{AB\cdots}~,
\end{equation}
and the differential equation for $\hat{L}_{AB}$ becomes a conservation type equation on the integrated operator.  However, note that only the transverse $x^{\perp}$ indices are summed.  Since $\mathcal{O}_{AB\cdots}$ must transform as a local operator in a full representation of the Lorentz group, this equation cannot be solved.\footnote{For instance, if $\mathcal{O}_{AB\cdots}$ were conserved on the index $B$, then we have 
\begin{equation}
\partial_{B}\mathcal{O}_{AB\cdots}\sim \partial_{-}\mathcal{O}_{A+\cdots}+\partial_{+}\mathcal{O}_{A-\cdots}~.
\end{equation}
The term involving $\partial_{-}$ above could plausibly be removed by the light-ray integral, however the term involving $\partial_{+}$ cannot vanish.  }

Thus we conclude that $\hat{L}_{AB}$ is zero and the commutator is 
\begin{align}\label{NE2}
[{\cal N}_A(x_1^\perp) ,  {\cal E}(x_2^\perp) ]  =  - i \delta^{(d-2)} (x_{12}^\perp) \, \partial_A{\cal E}(x_2^\perp)  
+i \partial_A \delta^{(d-2)} (x_{12}^\perp)  \,{\cal E}(x_2^\perp)\,.
\end{align}

\subsection*{$[{\cal N}_A, {\cal K}]$}

The commutator is constrained by causality and unitarity to take the form
\begin{align}
[{\cal N}_A(x_1^\perp) ,  {\cal K}(x_2^\perp)]  =
 i\delta^{(d-2)} (x_{12}^\perp) \,L_A(x_2^\perp)   
 +i \partial _B \delta^{(d-2)} (x_{12}^\perp) \,{L}_{AB}(x_2^\perp)\,,
\end{align}
where $L_A$ and $L_{AB}$ are again some non-local operators to be determined.  Integrating in $x_{1}^{\perp}$ and $x_{2}^{\perp}$ and enforcing Ward identities yields
\begin{equation}
L_{A}(x^{\perp})=-\partial_{A}\mathcal{K}(x^{\perp})~,\hspace{.5in}\partial_{A}\mathcal{K}(x^{\perp})=\partial_{B}L_{AB}~.
\end{equation}
The differential equation above is solved by writing $L_{AB}=\delta_{AB}\mathcal{K}+\hat{L}_{AB}$ where $\hat{L}_{AB}$ must now solve the homogenous equation $\partial_{B}\hat{L}_{AB}=0$.  As in the argument below \eqref{hatl1} this implies that $\hat{L}_{AB}$ vanishes and the commutator is
\begin{align}\label{NK2}
[{\cal N}_A(x_1^\perp) ,  {\cal K}(x_2^\perp)]  =
-  i \delta^{(d-2)}(x_{12}^\perp) \,  \partial _A {\cal K}(x_2^\perp)   + i \partial_A\delta^{(d-2)}(x_{12}^\perp)  \, {\cal K}(x_2^\perp)\,.
\end{align}

\subsection*{$[{\cal N}_A,{\cal N}_B]$}

Finally, let us consider the commutator between ${\cal N}_A$. The most general ansatz consistent with causality and the twist gap takes the form
\begin{align}\label{nnansatz}
&\left[ {\cal N}_A (x_1^\perp)  ,  {\cal N}_B(x_2^\perp) \right]\notag\\
=&i \delta^{(d-2)}(x_{12}^\perp) {L}_{AB}(x_2^\perp) + i\partial _{C}\delta^{(d-2)} (x_{12}^\perp) {L}_{ABC}(x_2^\perp)+ i\partial _{C}\partial _{D}\delta^{(d-2)} (x_{12}^\perp) {L}_{ABCD}(x_2^\perp)~,
\end{align}
where $L_{ABCD} = L_{ABDC}$.  
Integrating in $x_{1}^{\perp}$ and $x_{2}^{\perp}$ and enforcing the Ward identities implies that
\begin{eqnarray}\label{nnwards}
L_{AB}(x^{\perp})&=&-\partial_{A}\mathcal{N}_{B}(x^{\perp})~,\\
2\partial_{(A}\mathcal{N}_{B)}(x^{\perp})&=&\partial_{C}L_{ABC}(x^{\perp})+\partial_{C}\partial_{D}L_{ABCD}(x^{\perp})~. \nonumber
\end{eqnarray}
Next we enforce the constraint that the ansatz \eqref{nnansatz} is a consistent commutator.  This implies that it must be antisymmetric under the simultaneous exchange of 1 with 2 and $A$ with $B$.  After such an exchange, the proposed commutator \eqref{nnansatz} is
\begin{equation}\label{nnansatz1}
i \delta^{(d-2)}(x_{21}^\perp) {L}_{BA}(x_1^\perp) + i\partial _{C}\delta^{(d-2)} (x_{21}^\perp) {L}_{BAC}(x_1^\perp)+ i\partial _{C}\partial _{D}\delta^{(d-2)} (x_{21}^\perp) {L}_{BACD}(x_1^\perp)~.
\end{equation}
We can translate the operators above back to $x_{2}^{\perp}$ by writing $x_{1}^{\perp}=x_{2}^{\perp}+x_{12}^{\perp}$ and expanding in a Taylor series.  This expansion terminates at finite order since each term above is multiplied by a delta function and we consider $x^{n}\partial^{m}\delta(x)$ to be a trivial distribution if $n>m$.  Therefore \eqref{nnansatz1} becomes:
\begin{eqnarray}
&&i\delta^{(d-2)}(x_{21}^\perp) {L}_{BA}(x_2^\perp)+i\partial_{C}\delta^{(d-2)}(x_{21}^\perp) \left(L_{BAC}(x_{2}^{\perp})+x_{12}^{D}\partial_{D}L_{BAC}(x_{2}^{\perp})\right) \\
& +& i\partial_{C}\partial_{D}\delta^{(d-2)}(x_{21})\left(L_{BACD}(x_{2}^{\perp})+x_{12}^{E}\partial_{E}L_{BACD}(x_{2}^{\perp})+\frac{1}{2}x_{12}^{E}x_{12}^{F}\partial_{E}\partial_{F}L_{BACD}(x_{2}^{\perp})\right)~. \nonumber
\end{eqnarray}
The terms with explicit factors of $x_{12}^{\perp}$ above can now be integrated by parts since in any integral they must soak up the derivatives on the delta functions.  Further changing the arguments of the delta functions to $x_{12}^{\perp}$ (and being careful about signs), we deduce that in addition to \eqref{nnwards}, antisymmetry of the commutator requires
\begin{align}\label{commutatorconstraints}
&L_{ABCD}(x^{\perp})+L_{BACD}(x^{\perp})=0~, \hspace{.3in}L_{ABC}(x^{\perp})-L_{BAC}(x^{\perp})+\partial_{D}L_{ABCD}(x^{\perp})=0~.
\end{align}

By combining the equations \eqref{commutatorconstraints} and \eqref{nnwards} we can find simpler decoupled differential equations for $L_{ABCD}$ and $L_{ABC}$:
\begin{equation}\label{decupnn}
\partial_{C}\partial_{D}L_{ABCD}=0~, \hspace{.5in}\partial_{C}L_{ABC}=2\partial_{(A}\mathcal{N}_{B)}~.
\end{equation}
Using the argument below \eqref{hatl1}, the first equation above implies that $L_{ABCD}$ must vanish.  Note that here we must be slightly cautious about the fact that the differential operator acting on $L_{ABCD}$ is second order and therefore we need to exclude light-ray integrals of free fields.  However, $L_{ABCD}$ saturates the twist gap \eqref{twistgap}, and free fields always have smaller twist.  

Identical considerations allow us to fix $L_{ABC}$.  Indeed, decomposing as
\begin{equation}
L_{ABC}=\delta_{AC}\mathcal{N}_{B}+\delta_{BC}\mathcal{N}_{A}+\hat{L}_{ABC}~,
\end{equation}
we deduce from \eqref{decupnn} the homogeneous equation $\partial_{C}\hat{L}_{ABC}=0.$  Following the logic below \eqref{hatl1} we find that $\hat{L}_{ABC}$ vanishes.

In conclusion, the commutator is 
\begin{equation}\label{NN2}
\left[ {\cal N}_A (x_1^\perp)  ,  {\cal N}_B(x_2^\perp) \right]
= - i  \delta^{(d-2)}(x_{12}^\perp) \, \partial_A {\cal N}_B(x_2^\perp)   
+i \partial_A \delta^{(d-2)} (x_{12}^\perp ) \, {\cal N}_B(x_2^\perp) 
+i \partial_B \delta^{(d-2)} (x_{12}^\perp ) \, {\cal N}_A(x_2^\perp) ~.
\end{equation}

\subsection{Summary of the Algebra}

Collecting \eqref{EE2}, \eqref{KE2}, \eqref{KK2}, \eqref{NE2}, \eqref{NK2}, \eqref{NN2}, let us summarize the algebra of parallel  light-ray operators ${\cal E}$, $\cal K$, ${\cal N}_A$ on the light-sheet ${\cal S}$:
\begin{align}
\label{algebra}
&[{\cal E}(x^\perp_1) ,{\cal E}(x^\perp _2) ] =0\,,\notag\\
&[{\cal K}(x^\perp_1) ,{\cal K}(x^\perp _2) ] =0\,\notag\\
&[{\cal K}(x_1^\perp ) , {\cal E}(x_2^\perp)] = -i \delta^{(d-2)} (x_{12}^\perp) \,{\cal E}(x_2^\perp)\,,\notag\\
&[{\cal N}_A(x^\perp _1) , {\cal E}(x^\perp_2) ]  = - i \delta^{(d-2)}(x_{12}^\perp) \,\partial_A {\cal E}(x_2^\perp) +i \partial_A \delta^{(d-2)}(x_{12}^\perp )\, {\cal E}(x_2^\perp)\,,\\
&[{\cal N}_A(x^\perp_1)  , {\cal K}(x^\perp_2)  ] 
=  
-i \delta^{(d-2)} (x_{12}^\perp) \, \partial_A {\cal K}(x_2^\perp)
+i \partial_A \delta^{(d-2)} (x_{12}^\perp) \, {\cal K}(x_2^\perp) \,,\notag\\
&[{\cal N}_A(x^\perp_1) , {\cal N}_B(x_2^\perp)  ] =-i \delta^{(d-2)}(x_{12}^\perp)  \,  \partial_A {\cal N}_B(x_2^\perp)
+i \partial_{A} \delta^{(d-2)} (x_{12}^\perp) \,{\cal N}_{B}(x_2^\perp)
+i \partial_{B} \delta^{(d-2)} (x_{12}^\perp) \,{\cal N}_{A}(x_2^\perp)\,.\notag
\end{align}
The algebra of the smeared non-local operators on the light-sheet $\cal S$ (such as \eqref{smear}) can be straightforwardly derived from \eqref{algebra}.  In the following we will consider a subset of these smeared operators and identify their algebra as the generalized BMS algebra.

\subsection{The BMS Subalgebra}\label{sec:BMSsub}

The light-ray algebra \eqref{algebra} has an infinite-dimensional subalgebra that is identical to the generalized BMS algebra \cite{Campiglia:2014yka}.  Let $f(x^\perp)$ and $Y^A(x^\perp)$ be a scalar and a vector function in the traverse $\mathbb{R}^{d-2}$, respectively.   
Define
\begin{eqnarray}\label{BMSTR}
{\cal T}(f)  &\equiv& \int d^{d-2}x^\perp f(x^\perp) {\cal E}(x^\perp) ~,\\
{\cal R}(Y^A) &\equiv& \int d^{d-2}x^\perp Y^A(x^\perp) {\cal N}_A(x^\perp) 
+\frac {1}{d-2}   \int d^{d-2}x^\perp \partial_A Y^A(x^\perp)  {\cal K}(x^\perp)~.
\end{eqnarray}
The supertranslation ${\cal T}(f)$ is identified as a smeared version of the ANEC operator $\cal E$, while the superrotation ${\cal R}(Y^A)$ is a particular linear combination of smeared $\cal K$ and ${\cal N}_A$ light-ray operators.  Note that our identification is consistent with the geometric properties of the BMS algebra.  For instance, the supertranslation is an $x^\perp$-dependent translation along the $x^-$ direction, and precisely this transformation is generated by the smeared ANEC operator $\mathcal{T}(f)$.

Let us verify that the operators in \eqref{BMSTR} indeed generate the generalized BMS algebra \eqref{BMS}.  
First, since $[{\cal E}(x_1^\perp),  {\cal E}(x_2^\perp)]=0$, we immediately have $[{\cal T} (f_{1}) ,{\cal T}(f_{2})]=0$. Next, using \eqref{algebra}, we can easily compute 
\begin{align}
[ {\cal T}(f),{\cal R}(Y^A)  ] =i {\cal T} ( g)~,
\end{align}
where
\begin{align}
g=\frac {1}{d-2} ( \partial_A Y^A )f- Y^A\partial_A f \,.
\end{align}
Finally, let us check the commutator between the superrotations:
\begin{align}
&[{\cal R}(Y_1^A) , {\cal R}(Y_2^A) ] = i \int d^{d-2}x^\perp Y_3^A(x^\perp) {\cal N}_A(x^\perp) \notag\\
&
+\frac {i}{d-2}  \int d^{d-2}x_1^\perp \int dx_2^\perp  \,Y_1^A(x_1^\perp) \partial_BY_2^B(x_2^\perp)\left[-\delta^{(d-2)}( x_{12}^\perp) \partial_A {\cal K}(x_2^\perp) +\partial_A \delta^{(d-2)} (x_{12}^\perp) {\cal K}(x_2^\perp)\right]\notag\\
&
+\frac {i}{d-2}  \int d^{d-2}x_1^\perp \int d^{d-2}x_2^\perp
\partial_A Y_1^A(x_1^\perp) Y_2^B(x_2^\perp)
\left[ \delta^{(d-2)}( x_{21}^\perp) \partial_B {\cal K}(x_1^\perp) 
-\partial_B \delta^{(d-2)} (x_{21}^\perp) {\cal K}(x_1^\perp)\right] \notag\\
&= i \int d^{d-2}x^\perp Y_3^A(x^\perp) {\cal N}_A(x^\perp)
+\frac {i}{d-2}  \int d^{d-2}x^\perp (Y_1^A \partial _A\partial_B Y_2^B - Y_2^A \partial_A\partial_B Y_1^B) (x^\perp){\cal K}(x^\perp)
 \notag\\
 &= i {\cal R} (Y_3^A)\,,
\end{align}
where
\begin{align}
Y_3^A = Y_1^B (\partial_B Y_2^A)  -(\partial_B Y_1^A) Y_2^B\,.
\end{align}
These commutators exactly reproduce the generalized BMS algebra in $d$ dimensions \eqref{BMS}.

\section{Spin-One Conserved Currents}\label{sec:spinone}

The analysis of the previous section is easily extended to conformal field theories with a continuous global symmetry $G$.  Denote the associated spin-one conserved currents by $j_\mu^a(x)$, where $a$ is the adjoint index of the symmetry group $G$.  The charge operators are defined on a codimension one manifold $\Sigma$:
\begin{align}
Q^a = \int _\Sigma d\Sigma^\mu \, j_\mu^a(x)\,.
\end{align}
The commutator of the charge operators is
\begin{align}
[Q^a ,Q^b]  =if^{abc} Q^c\,,
\end{align}
where $f^{abc}$ are the structure constants of $G$. 

Define the light-ray operator 
\begin{align}
{\cal J}^a (x^\perp )  \equiv \int _{-\infty}^\infty dx^- \, j_-^a(x)\,,
\end{align}
on the light-sheet ${\cal S}:~x^+=0$.  If we further integrate in the $x^\perp$ directions with a constant profile, we reproduce the standard charge operators $Q^a$ defined on the light-sheet $\cal S$.   The light-ray operator ${\cal J}^a(x^\perp)$ has scaling dimension $\Delta= d-2$ and boost charge $J=0$.

Let us compute the commutator between these light-ray operators.  By causality and unitarity, the commutator must take the form
\begin{align}
[{\cal J}^a(x_1^\perp) , {\cal J}^b(x_2^\perp)]  =i \delta^{(d-2)}(x_{12}^\perp) L^{ab}(x_2^\perp) \,,
\end{align}
where terms with derivatives of $\delta^{(d-2)}(x_{12}^\perp)$ are ruled out by the twist gap bound.  Integrating both sides in $x_1^\perp$ and using the Ward identity of $Q^a$, we find $L^{ab}(x^\perp)  = f^{abc} {\cal J}^c(x^\perp)$.  Thus,
\begin{align}\label{JJ}
[{\cal J}^a(x_1^\perp) , {\cal J}^b(x_2^\perp)]  =i \delta^{(d-2)}(x_{12}^\perp) \, f^{abc} {\cal J}^c (x_2^\perp) \,.
\end{align}

We can smear the light-ray operator ${\cal J}^a(x^\perp)$ with an arbitrary function $g(x^\perp)$ to define a non-local operator supported on $\cal S$
\begin{align}
{\cal Q}^a(g) \equiv \int d^{d-2}x^\perp \, g(x^\perp)  {\cal J}^a(x^\perp)\,.
\end{align}
These operators have been considered in \cite{Strominger:2013lka} in the context of asymptotic symmetries of Yang-Mills theory. 
Using \eqref{JJ}, we find that ${\cal Q}^a(g)$  obeys the algebra
\begin{align}
[{\cal Q}^a(g_1 ) , {\cal Q}^b (g_2) ] =  i f^{abc}\, {\cal Q}^c (g_1g_2)\,.
\end{align}

\section{Light-ray Algebra in Correlation Functions}\label{sec:matrix}

In this section we would like to verify the light-ray algebra \eqref{algebra} inside correlation functions.  Specifically, we will compute the matrix elements of light-ray operators between two scalar primary operators ${\cal O}(x)$.  Such calculations were first considered in \cite{Hofman:2008ar}.

\subsection{Matrix Elements of One Light-ray Operator}

We start with the matrix element of a single light-ray operator between two scalar primaries ${\cal O}(x)$ of scaling dimension $\Delta$.  The case  of the ANEC operator was computed in \cite{Kravchuk:2018htv}.  Here we will set up the calculation and extend their results to other light-ray operators.

We will normalize our two-point function as
\begin{align}
\langle {\cal O}(x) {\cal O}(0) \rangle = { {\cal N} \over r^{d-2}}\,,
\end{align}
where $r^2 =x^\mu x_\mu$.  
The $\langle {\cal O}T{\cal O}\rangle$ three point function is given in (3.1) of \cite{Osborn:1993cr}:
\begin{align}\label{OTO}
\langle {\cal O}(x_2) T_{\mu\nu} (x_1) {\cal O}(x_3) \rangle
=a {1\over r_{12}^d r_{23}^{2\Delta-d} r_{13}^d}
  \left( 
  { X_{23\mu} X_{23\nu}  \over  X_{23}^2 }
   -\frac 1d \eta_{\mu\nu} \right)\,,
\end{align}
where
\begin{align}
&X_{23} ^\mu= {x_{12}^\mu \over r_{12}^2 }-{x_{13}^\mu \over r_{13}^2 }\,,
\end{align}
and 
\begin{align}
a \equiv -{\cal N}\Delta  {  d  \over (d-1)S_d}\,. 
\end{align}
Here $S_d= 2\pi^{d/2} /\Gamma(d/2)$.

We will place the energy-momentum tensor $T(x_1)$ at the light-sheet ${\cal S}$: $x_1^+=0$, while the two ${\cal O}$'s away from $\cal S$.  For such a configuration, we have
\begin{align}
\begin{split}
&r_{12}^2 =  - (x_{12}^- +i \epsilon) (- x_2^++ i\epsilon) + |x_{12}^\perp|^2\,,\\
&r_{13}^2 = - (x_{13}^- - i\epsilon)(- x_3^+ -i\epsilon)+|x_{13}^\perp|^2\,,\\
&r_{23}^2 = -x_{23}^+ x_{23}^- +|x_{23}^\perp|^2\,.
\end{split}
\end{align}
The $i\epsilon$'s are included to so that the operator ordering in Lorentzian signature is exactly as written in  \eqref{OTO}.

To simplify the calculation, we will restrict to $d=4$ for the rest of this subsection unless otherwise stated. 

Let us start with the three-point function involving the ANEC operator. The relevant component  is
\begin{align}\label{OTO2}
\langle {\cal O}(x_2) T_{--} (x_1){\cal O}(x_3) \rangle
=\frac a4
{1\over r_{12}^{2} r_{23}^{2\Delta-2} r_{13}^ {2}}
\left( {x_{12}^+ \over r_{12}^2 }  - {x_{13}^+ \over r_{13}^2 } \right)^2\,.
\end{align}
Since the three-point function \eqref{OTO2} falls off at large $x^-_1$ as $1/(x_1^-)^6$, 
the $x^-_1$ integral can be done by evaluating the residue where $r_{13}^2$ vanishes. This pole is located at
\begin{align}
x_1^-  = x_3^- -  {| x_{13}^\perp|^2\over x_3^+}  
+i \epsilon \left( 1+  { |x_{13}^\perp| ^2\over (x_3^+)^2} \right) 
+{\cal O}(\epsilon^2)\,.
\end{align}
Note that in particular this pole is on the upper half-plane.   We obtain
\begin{align}\label{OEO}
\langle {\cal O}(x_2)  {\cal E}(x_1^\perp) {\cal O}(x_3)\rangle
=-\frac {3a}{2}(2\pi i ) 
\,
 {1\over r_{23}^{2\Delta-2}}
{ (x_2^+ x_3^+)^2  
\over  \left[
x_2^+ ( - x_3^- x_3^+  + | x_{13}^\perp|^2 )
-  x_3^+  ( - x_2^- x_2^+ + |x_{12}^\perp|^2)
\right]^3} \,.
\end{align}
This is consistent with (3.21) of \cite{Kravchuk:2018htv}.  We also observe that this three-point function is well-defined and vanishes with the two ${\cal O}$'s approach the light-sheet $x^+=0$.

The matrix elements of ${\cal K}$  can be similarly computed to be
\begin{align}\label{OKO}
&\langle {\cal O}(x_2)  {\cal K}(x_1^\perp) {\cal O}(x_3)\rangle \notag\\
&
={3a\over 4}(2\pi i ) 
  {1\over r_{23}^{2\Delta-2}}
\,
{x_2^+ x_3^+ \left[
 x_2^+ (-  x_3^- x_3^+  +  |x_{13}^\perp|^2 )
 +x_3^+ ( -x_2^- x_2^+   +|x_{12}^\perp|^2)    
\right]
\over \left[
x_2^+ ( - x_3^- x_3^+  + | x_{13}^\perp|^2 )
-  x_3^+  ( - x_2^- x_2^+ + |x_{12}^\perp|^2)
\right]^3}  \,.
\end{align}

Finally, let us compute $\langle {\cal O}(x_2)  {\cal N}_A(x_1^\perp) {\cal O}(x_3)\rangle$. The relevant component of the three-point function \eqref{OTO} is
\begin{align}\label{OTO3}
\langle {\cal O}(x_2) T_{-A} (x_1){\cal O}(x_3) \rangle
=
-{a\over 2}
{1\over r_{12}^{2} r_{23}^{2\Delta-2} r_{13}^ {2}}
\left( {x_{12}^+ \over r_{12}^2 }  - {x_{13}^+ \over r_{13}^2 } \right)
\left(  {x_{12}^A \over r_{12}^2 }  - {x_{13}^A \over r_{13}^2 } \right)\,.
\end{align}
Since the three-point function \eqref{OTO2} falls off at large $x^-_1$ as $1/(x_1^-)^5$, 
the $x^-_1$ integral can again  be done  using the residue theorem.     We get
\begin{align}\label{ONO}
&\langle {\cal O}(x_2)  {\cal N}_A(x_1^\perp) {\cal O}(x_3)\rangle \notag\\
&
=-\frac {3a}{2}
(2\pi i ) 
  {1\over r_{23}^{2\Delta-2}}
\,
{x_2^+ x_3^+ \left(
x_{13}^A x_2^+ +x_{12}^A x_3^+ 
\right)
\over \left[
x_2^+ ( - x_3^- x_3^+  + | x_{13}^\perp|^2 )
-  x_3^+  ( - x_2^- x_2^+ + |x_{12}^\perp|^2)
\right]^3} \,.
\end{align}

\subsubsection{Subtleties on the Light-sheet}

So far we have assumed that the two scalar operators are away from the light-sheet ${\cal S}:~x^+=0$.  If we wish to extend our analysis to the case where the scalars are also placed on $\mathcal{S}$ we must confront the fact that matrix elements of light-ray operators are generally not well-defined when other local operators are on the same light-sheet.\footnote{We thank T. Hartman, P. Kravchuk, D. Simmons-Duffin, and D. Stanford for useful discussions on this point.}
 
  To illustrate this point, consider the limit where both scalars $\cal O$ tend to the the light-sheet $\cal S$ as:
\begin{align}\label{limit}
x_2^+ = \delta\,,~~~x_3^+  = \alpha\delta\,,~~~\delta\to 0\,,~~~~\alpha=\text{fixed}\,.
\end{align}
While the matrix element $\langle {\cal O}{\cal E}{\cal O}\rangle$ vanishes in this limit, the other two approach  $\alpha$-dependent constants:
\begin{align}
&\langle {\cal O}(x_2)  {\cal K}(x_1^\perp) {\cal O}(x_3)\rangle 
\to    -\frac{3a}{4} 
(2\pi i ) 
  {1\over r_{23}^{2\Delta-2}}
 { \alpha  (\alpha | x_{12}^\perp|^2 +|x_{13}^\perp|^2   )\over
  (\alpha |x_{12}^\perp|^2 -|x_{13}^\perp|^2)^3
  }\,,\\
 &  \langle {\cal O}(x_2)  {\cal N}_A(x_1^\perp) {\cal O}(x_3)\rangle\to 
 -3a(2\pi i ) 
  {1\over r_{23}^{2\Delta-2}}
 { \alpha (\alpha x_{12}^A +x_{13}^A )  \over (\alpha |x_{12}^\perp|^2 -|x_{13}^\perp|^2 )^3}\,.
\end{align}
Thus, unlike  $\langle {\cal O}{\cal E}{\cal O}\rangle$,  the matrix elements $\langle {\cal O} {\cal K}{\cal O}\rangle$ and  $\langle {\cal O} {\cal N}_A{\cal O}\rangle$ are not defined when every operator is on the same light-sheet.\footnote{Incidentally, the $x_1^-$ integral 
\begin{align}
\langle {\cal O}(x_2) \int dx_1^- (x_1^-)^n T_{--}(x_1) \,{\cal O}(x_3)\rangle
\end{align}
involving the light-ray operators considered in \cite{Casini:2017roe} is divergent for $n\ge 5$ and generic $x_2$, $x_3$.  For $4\ge n\ge2$, the matrix element is finite for generic $x_2$ and $x_3$, but 
 is singular in the limit \eqref{limit}.  On the the hand, it is a $\alpha$-dependent  constant when $n=1$ and vanishes  when $n=0$ as discussed above.}

In fact, a similar issue also appears for the light-ray operator ${\cal J}(x^\perp)$ of a spin-one conserved current studied in Section \ref{sec:spinone}.  The matrix element between two charged complex scalar primaries of dimension $\Delta$ in a $d$-dimensional CFT has been computed in (3.21) of \cite{Kravchuk:2018htv}:
\begin{align}
\langle {\cal O}(x_2) \, {\cal J}(x_2^\perp)\,{\cal O}^*(x_3)\rangle
= N\, {(x_2^+ x_3^+)^{d-2\over2} \over
\left[
x_2^+ ( - x_3^- x_3^+  + | x_{13}^\perp|^2 )
-  x_3^+  ( - x_2^- x_2^+ + |x_{12}^\perp|^2)
\right]^{d-2}
}\,,
\end{align}
where $N$ is a normalization constant.  Indeed, we see that, in the limit \eqref{limit}, this matrix element depends on the slope $\alpha$.

The subtleties above suggest that the matrix element of light-ray operators on the light-sheet generally requires a choice of regularization scheme.  For example, one can regulate the light-ray integrals by truncating the range of $x^{-}$ to be finite.\footnote{We thank T. Hartman for discussions on this point.}  For the remainder of this work, we will sidestep these complications by restricting our attention to correlation functions where the other local operators are not on the light-sheet $\mathcal{S}$.

\subsection{The Algebra of Light-ray Operators in Free Field Theory}

In this section we compute a matrix element of the commutator of light-ray operators and verify the algebra \eqref{algebra} in free scalar field theory.  In particular, we will verify $[ {\cal E}(x_1^\perp)  , {\cal E}(x_2^\perp) ] =0$ and $[ {\cal K}(x_1^\perp)  , {\cal E}(x_2^\perp) ] = - i \delta^{(d-2)} (x_{12}^\perp) \, {\cal E}(x_2^\perp)$.

 The computation  involves a four-point function  $\langle {\cal O} \,T_{--}T_{--}{\cal O}\rangle$.  We choose the operator $\mathcal{O}$ to be the free scalar field denoted as $\phi(x)$ with scaling dimension $\Delta={d-2\over2}$, normalized as in \cite{Osborn:1993cr}
\begin{align}\label{phiphi}
\langle \phi(x) \phi(0)\rangle = {1\over (d-2) S_d}  \,{1\over r^{d-2}}\,.
\end{align}

The energy-momentum tensor is 
\begin{align}
T_{\mu\nu}  =  \partial_\mu \phi \partial_\nu \phi -\frac 14 {1\over d-1} 
\left[ (d-2) \partial_\mu\partial_\nu +\eta_{\mu\nu} \partial^2 \right]\phi^2 \,.
\end{align}
The two-point function of the energy-momentum tensor generally takes the form
\begin{align}
&\langle T_{\mu\nu}(x) T_{\rho\sigma}(0)\rangle
=  {C_T \over r^{2d}  } \,\left[
\frac12 ( I_{\mu\rho}(x) I_{\nu\sigma} (x) +I_{\mu\sigma}(x) I_{\nu\rho}(x) )-\frac 1d \eta_{\mu\nu}\eta_{\rho\sigma}
\right]\,,\notag\\
&I_{\mu\nu}(x) =  \eta_{\mu\nu}  -2 {x_\mu x_\nu \over r^2}\,.
\end{align}
For a free scalar field, $C_T$ is
\begin{align}
C_T = {d\over d-1} {1\over (S_d)^2}\,.
\end{align}

The four-point function $\langle \phi(x_1) T_{--} (x_2) T_{--}(x_3)\phi(x_4)\rangle$ can be computed by Wick contractions:
\begin{align}\label{OTTO}
&\left( (d-2) S_d\right)^3  \times\, \langle \phi(x_1) T_{--} (x_2) T_{--}(x_3)\phi(x_4)\rangle \\
=& {d(d-2)^2\over 4(d-1)} {( x_{23}^+)^4 \over r_{14}^{d-2}  r_{23}^{2d+4}}\notag\\
+&4\left[\,
\left( {\partial\over \partial x_2^- }  {1\over r_{12}^{d-2} }\right)
\left(  {\partial^2\over \partial x_2^- \partial x_3^-}  {1\over r_{23}^{d-2} }\right)
\left( {\partial\over \partial x_3^- }  {1\over r_{34}^{d-2} }\right)
+(2\leftrightarrow 3)
\,\right]\notag\\
+&4
\left[\,    \left( {- (d-2)\over 4(d-1)}\right)^2 
\left(  {\partial^2 \over \partial ( x_2^-)^2 } {\partial^2 \over \partial ( x_3^-)^2 }  {1\over r_{12}^{d-2} r_{23}^{d-2} r_{34}^{d-2}  }\right) 
+(2\leftrightarrow 3)\, \right]
\notag\\
+&4\left\{  
\left[ \,
 \left( {- (d-2)\over 4(d-1)}\right)
{\partial^2 \over \partial ( x_2^-)^2 } \left(
 {1\over r_{12}^{d-2}} \left( 
 {\partial \over \partial  x_3^- }  {1\over  r_{23}^{d-2} }\right)
 \left(
 {\partial \over \partial  x_3^- }  {1\over  r_{34}^{d-2} }\right)\right)
 +(2\leftrightarrow 3) \,
 \right]
 +(1\leftrightarrow 4) \,\right\} \,. \notag
\end{align}
The second line is the disconnected contribution $\langle T_{--}(x_2)T_{--} (x_3)\rangle \langle\phi(x_1)\phi(x_4)\rangle$.  The overall factor $((d-2)S_d)^3$ comes from our normalization of the scalar two-point function \eqref{phiphi} and appears on the left-hand side above.  The Lorentzian separation is $r_{ij}^2 = -x_{ij}^+ x_{ij}^- +|x_{ij}^\perp|^2 $.

To enforce the appropriate time ordering we use the following $i\epsilon$ prescription:
\begin{align}
x_1^\pm \to x_1^\pm -2 i\epsilon\,,~~~
x_2^\pm \to x_2^\pm - i\epsilon\,,~~~
x_3^\pm \to x_3^\pm + i\epsilon\,,~~~
x_4^\pm \to x_4^\pm +2 i\epsilon\,.
\end{align}
Furthermore, we place the two energy-momentum tensors on the light-sheet $\cal S$, i.e.\ 
\begin{align}
\text{Re}(x_2^+ )=\text{Re}(x_3^+ )=0\,.
\end{align}
Below, we further restrict to $d=4.$

\subsection*{$\langle \phi [{\cal E},{\cal E}] \phi\rangle$}

To obtain the ANEC operator, we first integrate \eqref{OTTO} in $x_2^-$.  The integral can be done by deforming the contour to circle the pole where $r_{12}^2=0$.  This pole is located at
\begin{align}
x_2^- =   x_1^- - {|x_{12}^\perp|^2 \over x_1^+ } - i \epsilon \left(1+ {|x_{12}^\perp|^2\over (x_1^+)^2}\right) +{\cal O}(\epsilon^2)\,.
\end{align}
It turns out that only the first term on the third line of \eqref{OTTO} contributes.  In this way we obtain $\langle \phi(x_1) {\cal E}(x_2^\perp) T_{--}(x_3)\phi(x_4)\rangle$.  Next we integrate in $x_3^-$ and deform the contour to circle the pole where $r_{34}^2=0$. This pole is located at 
\begin{align}
x_3^-  = x_4^- - {|x_{34}^\perp|^2\over x_4^+ } + i\epsilon \left(1+ {|x_{34}^\perp|^2 \over (x_4^+)^2}\right)+{\cal O}(\epsilon^2)\,.
\end{align}

After a somewhat tedious calculation, we obtain
\begin{align}
\langle \phi(x_1) {\cal E}(x_2^\perp) {\cal E}(x_3^\perp) \phi(x_4)\rangle
= {96\over \pi^4}
  \,   {
\epsilon^4   \over
x_1^+ x_4^+(  |x_{23}^\perp|^2 + i\epsilon A )^5
}\,,
\end{align}
where $A= {1\over x_1^+ x_4^+}\left( 
2 x_{14}^- x_1^+ x_4^+   -  2x_{4}^+ |x_{12}^\perp|^2 
+2 x_1^+ |x_{34}^\perp|^2  +x_{14}^+|x_{23}^\perp|^2\right)$.  This result should be interpreted as a distribution: it vanishes when $x_{23}^\perp$ is non-zero, but diverges as $\epsilon\rightarrow 0$ when $x_{23}^\perp$ vanishes.  By inspection, we see that  the four-point function is symmetric in $x_2^\perp$ and $x_3^\perp$, and hence we find 
\begin{align}
\langle \phi(x_1) [{\cal E}(x_2^\perp) ,{\cal E}(x_3^\perp)] \phi(x_4)\rangle=0\,,
\end{align} 
which is consistent with the light-ray algebra $[{\cal E} ,{\cal E}]=0$.  This matrix element is also studied in \cite{KKSDZ}.

\subsection*{$\langle \phi [{\cal K},{\cal E}] \phi\rangle$}

Finally, we consider the more non-trivial light-ray commutator $ [{\cal K}(x_1^\perp),{\cal E}(x_2^\perp) ] = - i \delta^{(2)} (x_{12}^\perp) \, {\cal E}(x_2^\perp)$.  The four-point functions $ \langle \phi(x_1) {\cal K}(x_2^\perp) {\cal E}(x_3^\perp) \phi(x_4)\rangle$ and $ \langle \phi(x_1) {\cal E}(x_2^\perp) {\cal K}(x_3^\perp) \phi(x_4)\rangle$  can be straightforwardly computed following the previous steps  by inserting $x^-$ into the integral.  Again only the first term in the third line of \eqref{OTTO} contributes. The final result is
\begin{align}
\langle \phi(x_1) [{\cal K}(x_2^\perp) , {\cal E}(x_3^\perp) ]\phi(x_4)\rangle
= 
{24 i \over\pi^4}  \,
{\epsilon^3
( |x_{23}^\perp|^2 - i\epsilon B)
\over 
x_1^+ x_4^+( |x_{23}^\perp|^2 + i \epsilon A)^5
}\,,
\end{align}
where 
$B={1\over x_1^+x_4^+}  \left(2x_{14}^- x_1^+ x_4^+  - 2x_4^+ |x_{12}^\perp|^2  +2x_1^+ |x_{34}^\perp|^2  - x_{14}^+ |x_{23}^\perp|^2\right)$.  
We want to show that 
this matrix element $\langle \phi(x_1)[{\cal K}(x_2^\perp) , {\cal E}(x_3^\perp) ]\phi(x_4)\rangle$ of the commutator is proportional to the transverse delta function $\delta^{(2)}(x_{23}^\perp)$. Indeed, the matrix element 
has the following properties:
\begin{itemize}
\item If $x_{23}^\perp \neq0$,  it is 0 as $\epsilon\to0$.
\item If $x_{23}^\perp = 0$,  it is singular as $\epsilon\to0$.
\end{itemize}
To finish the argument, we need to make sure upon integration in $x_2^\perp$, 
\begin{align}
\int d^2 x_2^\perp \,
\langle \phi(x_1) [{\cal K}(x_2^\perp) , {\cal E}(x_3^\perp) ]\phi(x_4)\rangle= -i \langle \phi(x_1) {\cal E}(x_3^\perp)  \phi(x_4)\rangle\,.
\end{align}
  We will check this in a special case where $x_1^\perp =x_3^\perp = 0$.  In this case the transverse integral in $x_2^\perp$ can be done analytically:
\begin{align}
\int d^2 x_2^\perp \, \langle \phi(x_1) [{\cal K}(x_2^\perp) , {\cal E}(x_3^\perp) ]\phi(x_4)\rangle
= {1\over 2\pi^3} {( x_4^+)^2 \over x_1^+ (|x_4^\perp|^2 +x_{14}^- x_4^+)^3}
\,,~~~~x_1^\perp =x_3^\perp =0\,.
\end{align}
The right-hand side is indeed the three-point function $-i \langle \phi(x_1) {\cal E}(x_3^\perp)  \phi(x_4)\rangle$ computed in \eqref{OEO} with $\Delta=1$, $d=4$,  $a= -{1\over 6\pi^4} $, and $x_1^\perp = x_3^\perp=0$.  Moreover, if we integrate the four-point function $\langle \phi(x_1) [{\cal K}(x_2^\perp) , {\cal E}(x_3^\perp) ]\phi(x_4)\rangle$ against a positive power of $x_2^\perp$, the answer vanishes as $\epsilon\to0$. This is consistent with the fact that there is no derivatives of the delta function in $[{\cal K}(x_2^\perp) , {\cal E}(x_3^\perp) ]$.

\section*{Acknowledgements}

We are grateful to N. Afkhami-Jeddi, T. Hartman, D. Kapec, P. Kravchuk,  J. Maldacena,  P. Mitra, D. Simmons-Duffin, D. Stanford, E. Witten for insightful discussions.  We also thank T. Hartman, P. Kravchuk, J. Maldacena, and A. Strominger for comments on a draft.
We thank the Bootstrap 2018 workshop at  Caltech for its hospitality during the course of this work. 
C.C. is supported by DOE grant de-sc0009988. 
S.H.S. is supported by the National Science Foundation grant PHY-1606531 and the Roger Dashen Membership.

\appendix

\section{Conformal Ward Identities for Light-ray Operators}\label{app:ward}

In Section \ref{sec:alg} we have used the Ward identities of the translation and the boost symmetry to constrain the algebra of the light-ray operators \eqref{algebra}. 
In this appendix we will check that \eqref{algebra} is also consistent with the Ward identities of some other generators of the conformal algebra.

On any codimension one manifold $\Sigma$, the global charges for the translation $P_\mu$,  the Lorentz transformation $J_{\mu\nu}$, the dilation $D$, and the special conformal transformation $K_\mu$  are
\begin{align}
\begin{split}
&P_\mu = \int d\Sigma^\rho \, T_{\rho\mu}\,,~~~~~
J_{\mu\nu} =2  \int_{\Sigma}d\Sigma^\rho \, x_{[\mu} T_{\nu] \rho}(x)\,, ~~~\\
&D = \int_{\Sigma}d\Sigma^\rho \, x^\mu T_{\rho\mu}\,,~~~
K_\mu  = \int_{\Sigma}d\Sigma^\rho \,\left(
2x_\mu x^\nu T_{\rho\nu}  - x^2 T_{\rho\mu}
\right)\,.
\end{split}
\end{align}
The action of these global charges on a local operator ${\cal O}(x)$ of dimension $\Delta$ is  (here we suppress the spin indices of ${\cal O}$) 
\begin{align}
&[ P_\mu , {\cal O}(x)] = -i \partial_\mu {\cal O}(x)\,,\label{Pward}\\
&[ J_{\mu\nu} ,   {\cal O}(x) ] = -2ix_{[\mu}  \partial_{\nu] } {\cal O}(x)+ M_{\mu\nu} \cdot {\cal O}(x) \,,\label{Jward}\\
&[D , {\cal O}(x)] =   -  i x^\mu \partial_\mu {\cal O}(x)- i  \Delta{\cal O}(x) \,,\label{Dward}\\
&[K_\mu  , {\cal O}(x) ] = -2 i x_\mu x^\nu \partial_\nu  {\cal O}(x)+i x^2 \partial_\mu {\cal O}(x) 
-2i\Delta x_\mu {\cal O}(x)
 +2x^\nu M_{\mu\nu} \cdot {\cal O}(x)\,,\label{Kward}
\end{align}
where  $M_{\mu\nu}$, for any fixed pair of $\mu\nu$,  is a finite-dimensional matrix in the Lorentz  representation of  ${\cal O}(x)$. For example, for a vector field $V_\mu(x)$, we have $M_{\mu\nu}\cdot V_\rho(x) = (M_{\mu\nu})_\rho ^\sigma V_\sigma (x)=  -i (\eta_{\mu\rho}V_\nu(x)  - \eta_{\nu \rho} V_\mu(x))$.  The action of the global charges on light-ray operators is given by integrating the above along $x^{-}$.

We will focus on the transverse rotation $J_{AB}$, the dilation $D$, and the transverse special conformal transformation $K_A$. On the light-sheet ${\cal S}:~x^+=0$, they take the form
\begin{align}
&J_{AB} =2  \int d^{d-2}x^\perp \, x_{[A } {\cal N}_{B]}(x^\perp)\,,\\
&D=  \int d^{d-2} x^\perp \,  \left(\, {\cal K}(x^\perp)  +x_A \, {\cal N}_A(x^\perp) \,\right)\,,\\
&K_A = \int d^{d-2}x^\perp \, \left(\,
2x_A {\cal K}(x^\perp) + 2x_Ax_B {\cal N}_B (x^\perp) - |x^\perp|^2 {\cal N}_A(x^\perp)
\,\right)\,.
\end{align}

Let us start with the commutator $[ J_{CA}  ,  {\cal E}(x_2^\perp) ] $:
\begin{align}
[ J_{CA}  ,  {\cal E}(x^\perp) ] 
 = 2\int d^{d-2}x_1^\perp \, x_{1[C} [ {\cal N}_{A]} (x_1^\perp)  ,  {\cal E}(x^\perp)]
=  -2i x_{[C} \partial_{A]}{\cal E}(x^\perp)\,,
\end{align}
which is consistent with \eqref{Jward}.   Let us move on to $[D,{\cal E}(x^\perp)]$:
\begin{align}
[D,  {\cal E}(x^\perp) ] 
&=\int d^{d-2}x_1^\perp \, \left[ \,
{\cal K}(x_1^\perp )  +x_{1A} {\cal N}_A(x_1^\perp), {\cal E}(x^\perp) 
\,\right] \notag\\
&
= - i  x_A \partial_A {\cal E}(x^\perp) - i (d-1 ){\cal E}(x^\perp)\,,
\end{align}
which is consistent with \eqref{Dward} with $\Delta=d-1$.  Recall that ${\cal E}(x^\perp)$ is placed on the light-sheet ${\cal S}$ where $x^+=0$.  Moreover, $\partial_- {\cal E}=0$.  Moving on to $[K_A, {\cal E}(x^\perp)]$:
\begin{align}
[K_A ,{\cal E}(x^\perp) ] 
&=\int d^{d-2}x_1^\perp \, \left[\, 
2x_{1A} {\cal K}(x_1^\perp)+ ( 2x_{1A} x_{1B} -\delta_{AB}|x_1^\perp|^2 ) {\cal N}_B(x^\perp_1)  , {\cal E}(x^\perp)
\,\right]\notag\\
&=   - 2i (x_Ax_B - \delta_{AB}|x^\perp|^2)\partial_A{\cal E}(x^\perp)- 2i (d-1)x_A {\cal E}(x^\perp)
\end{align}
which agrees with \eqref{Kward}.   Similarly the commutators with $\cal K$ can be shown to be consistent with \eqref{Jward}, \eqref{Dward}, and \eqref{Kward}.

Let us now check the commutators involving ${\cal N}_A$.  We have
\begin{align}
[J_{CA} , {\cal N}_B(x^\perp) ] 
=  - i x_C \partial_A {\cal N}_B(x^\perp) 
+ i x_A \partial_C {\cal N}_B(x^\perp) 
-i  \delta_{BC} {\cal N}_A(x^\perp) +i  \delta_{BA} {\cal N}_C(x^\perp) \,,
\end{align}
which is consistent \eqref{Jward}.  Next,
\begin{align}
[D, {\cal N}_B(x^\perp)]   &=\int d^{d-2}x_1^\perp \, \left[ \,
{\cal K}(x_1^\perp )  +x_{1A} {\cal N}_A(x_1^\perp), {\cal N}_B(x^\perp) 
\,\right] \notag\\
&=
-ix_A \partial_A{\cal N}_B(x^\perp)
-  i (d-1)  {\cal N}_B(x^\perp)  \,.
\end{align}
Finally, 
\begin{align}
&[K_A,{\cal N}_B(x_2^\perp)  ] =
\int d^{d-2}x_1^\perp \, \left[\, 
2x_{1A} {\cal K}(x_1^\perp)+ ( 2x_{1A} x_{1C} -\delta_{AC}|x_1^\perp|^2 ) {\cal N}_C(x^\perp_1)  , {\cal N}_B(x^\perp_2)
\,\right]\notag\\
&=  2i  \int d^{d-2}x_1^\perp \,x_{1A}\left (\delta^{(d-2)} (x_{21}) \partial_B{\cal K}(x_1^\perp)  - \partial_B\delta^{(d-2)}(x_{21}^\perp)    {\cal K}(x_1^\perp)\,     \right)\notag\\
&
+ \int d^{d-2} x_1^\perp \, (2x_{1A} x_{1C} -\delta_{AC} |x_1^\perp|^2)
\left(\,
-i \delta^{(d-2)}(x_{12}^\perp) \partial_C{\cal N}_B(x_2^\perp)
+2i \partial_{(C} \delta^{(d-2)}(x_{12}^\perp)  {\cal N}_{B)}(x_2^\perp)\,
\right) \notag\\
&=
- i ( 2x_{2A} x_{2C} -\delta_{AC}|x_2^\perp|^2 ) \partial_C {\cal N}_B(x_2^\perp)
-2i  (d-1) x_{2A} {\cal N}_B(x_2^\perp)\notag\\
&
- 2 i  ( \delta_{AB} x_{2C}    -  \delta_{AC} x_{2B} ) {\cal N}_C(x_2^\perp)
-2i \delta_{AB} {\cal K}(x_2^\perp)
\,,
\end{align}
which agrees with \eqref{Jward} acting on the integrand of ${\cal N}_B$.  Here we have used 
\begin{align}
&2\int_{-\infty}^\infty dx^- \, x^\nu M_{A\nu}\cdot T_{-B}(x) 
=  2 \int_{-\infty }^\infty dx^-  \, x_C  M_{AC} \cdot  T_{-B}(x)
+  2\int_{-\infty }^\infty dx^-  \,  x^-  M_{A-} \cdot  T_{-B}(x)\notag\\
=&- 2ix_C (\delta_{AB} {\cal N}_C  - \delta_{CB}{\cal N}_A)
- 2i \delta_{AB} {\cal K}\,.
\end{align}

\bibliography{Lightraydraft}

\providecommand{\href}[2]{#2}\begingroup\raggedright\begin{thebibliography}{10}

\bibitem{Bondi:1962px}
H.~Bondi, M.~G.~J. van~der Burg, and A.~W.~K. Metzner, ``{Gravitational waves
  in general relativity. 7. Waves from axisymmetric isolated systems},'' {\em
  Proc. Roy. Soc. Lond.} {\bf A269} (1962)
21--52.
%%CITATION = PRSLA,A269,21;%%.

\bibitem{Sachs:1962wk}
R.~K. Sachs, ``{Gravitational waves in general relativity. 8. Waves in
  asymptotically flat space-times},'' {\em Proc. Roy. Soc. Lond.} {\bf A270}
  (1962)
103--126.
%%CITATION = PRSLA,A270,103;%%.

\bibitem{Sachs:1962zza}
R.~Sachs, ``{Asymptotic symmetries in gravitational theory},'' {\em Phys. Rev.}
  {\bf 128} (1962)
2851--2864.
%%CITATION = PHRVA,128,2851;%%.

\bibitem{Gaiotto:2014kfa}
D.~Gaiotto, A.~Kapustin, N.~Seiberg, and B.~Willett, ``{Generalized Global
  Symmetries},'' {\em JHEP} {\bf 02} (2015) 172,
\href{http://www.arXiv.org/abs/1412.5148}{{\tt 1412.5148}}.
%%CITATION = ARXIV:1412.5148;%%.

\bibitem{Casini:2017roe}
H.~Casini, E.~Teste, and G.~Torroba, ``{Modular Hamiltonians on the null plane
  and the Markov property of the vacuum state},'' {\em J. Phys.} {\bf A50}
  (2017), no.~36, 364001,
\href{http://www.arXiv.org/abs/1703.10656}{{\tt 1703.10656}}.
%%CITATION = ARXIV:1703.10656;%%.

\bibitem{Balakrishnan:2017bjg}
S.~Balakrishnan, T.~Faulkner, Z.~U. Khandker, and H.~Wang, ``{A General Proof
  of the Quantum Null Energy Condition},''
\href{http://www.arXiv.org/abs/1706.09432}{{\tt 1706.09432}}.
%%CITATION = ARXIV:1706.09432;%%.

\bibitem{Hartman:2016lgu}
T.~Hartman, S.~Kundu, and A.~Tajdini, ``{Averaged Null Energy Condition from
  Causality},'' {\em JHEP} {\bf 07} (2017) 066,
\href{http://www.arXiv.org/abs/1610.05308}{{\tt 1610.05308}}.
%%CITATION = ARXIV:1610.05308;%%.

\bibitem{Faulkner:2016mzt}
T.~Faulkner, R.~G. Leigh, O.~Parrikar, and H.~Wang, ``{Modular Hamiltonians for
  Deformed Half-Spaces and the Averaged Null Energy Condition},'' {\em JHEP}
  {\bf 09} (2016) 038,
\href{http://www.arXiv.org/abs/1605.08072}{{\tt 1605.08072}}.
%%CITATION = ARXIV:1605.08072;%%.

\bibitem{Hofman:2008ar}
D.~M. Hofman and J.~Maldacena, ``{Conformal collider physics: Energy and charge
  correlations},'' {\em JHEP} {\bf 05} (2008) 012,
\href{http://www.arXiv.org/abs/0803.1467}{{\tt 0803.1467}}.
%%CITATION = ARXIV:0803.1467;%%.

\bibitem{Kravchuk:2018htv}
P.~Kravchuk and D.~Simmons-Duffin, ``{Light-ray operators in conformal field
  theory},''
\href{http://www.arXiv.org/abs/1805.00098}{{\tt 1805.00098}}.
%%CITATION = ARXIV:1805.00098;%%.

\bibitem{KKSDZ}
M.~Kologlu, P.~Kravchuk, D.~Simmons-Duffin, and A.~Zhiboedov$, $ to~appear.

\bibitem{Strominger:2013jfa}
A.~Strominger, ``{On BMS Invariance of Gravitational Scattering},'' {\em JHEP}
  {\bf 07} (2014) 152,
\href{http://www.arXiv.org/abs/1312.2229}{{\tt 1312.2229}}.
%%CITATION = ARXIV:1312.2229;%%.

\bibitem{He:2014laa}
T.~He, V.~Lysov, P.~Mitra, and A.~Strominger, ``{BMS supertranslations and
  Weinberg's soft graviton theorem},'' {\em JHEP} {\bf 05} (2015) 151,
\href{http://www.arXiv.org/abs/1401.7026}{{\tt 1401.7026}}.
%%CITATION = ARXIV:1401.7026;%%.

\bibitem{Strominger:2017zoo}
A.~Strominger, ``{Lectures on the Infrared Structure of Gravity and Gauge
  Theory},''
\href{http://www.arXiv.org/abs/1703.05448}{{\tt 1703.05448}}.
%%CITATION = ARXIV:1703.05448;%%.

\bibitem{Weinberg:1965nx}
S.~Weinberg, ``{Infrared photons and gravitons},'' {\em Phys. Rev.} {\bf 140}
  (1965)
B516--B524.
%%CITATION = PHRVA,140,B516;%%.

\bibitem{Kapec:2016aqd}
D.~Kapec, A.-M. Raclariu, and A.~Strominger, ``{Area, Entanglement Entropy and
  Supertranslations at Null Infinity},'' {\em Class. Quant. Grav.} {\bf 34}
  (2017), no.~16, 165007,
\href{http://www.arXiv.org/abs/1603.07706}{{\tt 1603.07706}}.
%%CITATION = ARXIV:1603.07706;%%.

\bibitem{Strominger:2014pwa}
A.~Strominger and A.~Zhiboedov, ``{Gravitational Memory, BMS Supertranslations
  and Soft Theorems},'' {\em JHEP} {\bf 01} (2016) 086,
\href{http://www.arXiv.org/abs/1411.5745}{{\tt 1411.5745}}.
%%CITATION = ARXIV:1411.5745;%%.

\bibitem{Pasterski:2015tva}
S.~Pasterski, A.~Strominger, and A.~Zhiboedov, ``{New Gravitational
  Memories},'' {\em JHEP} {\bf 12} (2016) 053,
\href{http://www.arXiv.org/abs/1502.06120}{{\tt 1502.06120}}.
%%CITATION = ARXIV:1502.06120;%%.

\bibitem{Barnich:2011ct}
G.~Barnich and C.~Troessaert, ``{Supertranslations call for superrotations},''
  {\em PoS} {\bf CNCFG2010} (2010) 010,
  \href{http://www.arXiv.org/abs/1102.4632}{{\tt 1102.4632}}.
[Ann. U. Craiova Phys.21,S11(2011)].
%%CITATION = ARXIV:1102.4632;%%.

\bibitem{Campiglia:2014yka}
M.~Campiglia and A.~Laddha, ``{Asymptotic symmetries and subleading soft
  graviton theorem},'' {\em Phys. Rev.} {\bf D90} (2014), no.~12, 124028,
\href{http://www.arXiv.org/abs/1408.2228}{{\tt 1408.2228}}.
%%CITATION = ARXIV:1408.2228;%%.

\bibitem{Banks:2003vp}
T.~Banks, ``{A Critique of pure string theory: Heterodox opinions of diverse
  dimensions},''
\href{http://www.arXiv.org/abs/hep-th/0306074}{{\tt hep-th/0306074}}.
%%CITATION = HEP-TH/0306074;%%.

\bibitem{Barnich:2009se}
G.~Barnich and C.~Troessaert, ``{Symmetries of asymptotically flat 4
  dimensional spacetimes at null infinity revisited},'' {\em Phys. Rev. Lett.}
  {\bf 105} (2010) 111103,
\href{http://www.arXiv.org/abs/0909.2617}{{\tt 0909.2617}}.
%%CITATION = ARXIV:0909.2617;%%.

\bibitem{Barnich:2010eb}
G.~Barnich and C.~Troessaert, ``{Aspects of the BMS/CFT correspondence},'' {\em
  JHEP} {\bf 05} (2010) 062,
\href{http://www.arXiv.org/abs/1001.1541}{{\tt 1001.1541}}.
%%CITATION = ARXIV:1001.1541;%%.

\bibitem{Barnich:2011mi}
G.~Barnich and C.~Troessaert, ``{BMS charge algebra},'' {\em JHEP} {\bf 12}
  (2011) 105,
\href{http://www.arXiv.org/abs/1106.0213}{{\tt 1106.0213}}.
%%CITATION = ARXIV:1106.0213;%%.

\bibitem{Mack:1975je}
G.~Mack, ``{All unitary ray representations of the conformal group SU(2,2) with
  positive energy},'' {\em Commun. Math. Phys.} {\bf 55} (1977)
1.
%%CITATION = CMPHA,55,1;%%.

\bibitem{doi:10.1063/1.1705183}
N.~T. Evans, ``Discrete series for the universal covering group of the 3 + 2 de
  sitter group,'' {\em Journal of Mathematical Physics} {\bf 8} (1967), no.~2,
  170--184,
  \href{http://www.arXiv.org/abs/https://doi.org/10.1063/1.1705183}{{\tt
  https://doi.org/10.1063/1.1705183}}.

\bibitem{Minwalla:1997ka}
S.~Minwalla, ``{Restrictions imposed by superconformal invariance on quantum
  field theories},'' {\em Adv. Theor. Math. Phys.} {\bf 2} (1998) 783--851,
\href{http://www.arXiv.org/abs/hep-th/9712074}{{\tt hep-th/9712074}}.
%%CITATION = HEP-TH/9712074;%%.

\bibitem{Cordova:2017dhq}
C.~C\'{o}rdova and K.~Diab, ``{Universal Bounds on Operator Dimensions from the
  Average Null Energy Condition},'' {\em JHEP} {\bf 02} (2018) 131,
\href{http://www.arXiv.org/abs/1712.01089}{{\tt 1712.01089}}.
%%CITATION = ARXIV:1712.01089;%%.

\bibitem{Strominger:2013lka}
A.~Strominger, ``{Asymptotic Symmetries of Yang-Mills Theory},'' {\em JHEP}
  {\bf 07} (2014) 151,
\href{http://www.arXiv.org/abs/1308.0589}{{\tt 1308.0589}}.
%%CITATION = ARXIV:1308.0589;%%.

\bibitem{Osborn:1993cr}
H.~Osborn and A.~C. Petkou, ``{Implications of conformal invariance in field
  theories for general dimensions},'' {\em Annals Phys.} {\bf 231} (1994)
  311--362,
\href{http://www.arXiv.org/abs/hep-th/9307010}{{\tt hep-th/9307010}}.
%%CITATION = HEP-TH/9307010;%%.

\end{thebibliography}\endgroup
\bibliographystyle{utphys}

\end{document}